\documentclass{emulateapj}
\usepackage{natbib}
\usepackage{epsfig}
\usepackage{graphicx}
\usepackage{subfigure}
\usepackage{float}
\usepackage{amsmath}
\usepackage{color}
\usepackage{amssymb}
\usepackage{amsfonts}
\usepackage[colorlinks,linkcolor=blue,anchorcolor=green,citecolor=blue]{hyperref}
\usepackage{amssymb}
\usepackage{upgreek}
\usepackage[normalem]{ulem}
\usepackage{verbatim}

\def\be{\begin{equation}}
\def\ee{\end{equation}}

\shorttitle{Time-Frequency Downward Drifting of Repeating FRB}
\shortauthors{Wang et al.}
\begin{document}

\title{On the Time-Frequency Downward Drifting of Repeating Fast Radio Bursts}
\author{Weiyang Wang\altaffilmark{1,2,3}, Bing Zhang\altaffilmark{4}, Xuelei Chen\altaffilmark{1,2,7}, Renxin Xu\altaffilmark{3,5,6}}\email{wywang@bao.ac.cn}
\affil{$^1$Key Laboratory for Computational Astrophysics, National Astronomical Observatories, Chinese Academy of Sciences, 20A Datun Road, Beijing 100101, China}
\affil{$^2$University of Chinese Academy of Sciences, Beijing 100049, China}
\affil{$^3$School of Physics and State Key Laboratory of Nuclear Physics and Technology, Peking University, Beijing 100871, China}
\affil{$^4$Department of Physics and Astronomy, University of Nevada, Las Vegas, NV 89154, USA}\email{zhang@physics.unlv.edu}
\affil{$^5$Kavli Institute for Astronomy and Astrophysics, Peking University, Beijing 100871, China} 
\affil{$^6$Department of Astronomy, School of Physics, Peking University, Beijing 100871, China}
\affil{$^7$Center for High Energy Physics, Peking University, Beijing 100871, China}
\begin{abstract}
The newly discovered second repeating fast radio burst (FRB) source, FRB 180814.J0422+73, was reported to exhibit a time-frequency downward drifting pattern, which is also seen in the first repeater FRB 121102.
We propose a generic geometrical model to account for the observed downward drifting of sub-pulse frequency, 
within the framework of coherent curvature radiation by bunches of electron-positron pairs in the magnetosphere of a neutron star.
A sudden trigger event excites these coherent bunches of charged particles, which stream outwards along open field lines. 
As the field lines sweep across the line of sight, the bunches seen later have traveled farther into the 
less curved part of the magnetic field lines, thus emitting at lower frequencies.
We use this model to explain the time-frequency downward drifting in two FRB generation scenarios, 
the transient pulsar-like sparking from the inner gap region of a slowly rotating neutron star, and the 
externally-triggered magnetosphere reconfiguration known as the ``cosmic comb''.
\end{abstract}


\section{Introduction}

Fast radio bursts (FRBs) are mysterious millisecond-duration astronomical radio transients with large dispersion measures in excess of the Galactic value (DM$\gtrsim200\,{\rm pc\,cm^{-3}}$, \citealt{Lorimer07,Keane12,Thornton13,Kulkarni14,Petroff15,Petroff16,Chatterjee17}).
The cosmological origin of FRBs was established after FRB 121102, the first repeating source \citep{Spitler16}, was localized in a star-forming dwarf galaxy at $z = 0.193$ with an associated persistent radio source \citep{Bassa17,Chatterjee17,Marcote17,Tendulkar17} and an extreme magneto-ionic environment \citep{Michilli18}.

Recently, the \cite{Chime19} reported the discovery of the second repeating FRB source, FRB 180814.J0422+73.
Very intriguingly, both FRB 121102 and FRB 180814.J0422+73 showed an interesting sub-pulse time-frequency downward drifting pattern in at least some of their bursts. For these bursts, each burst have several sub-pulses, with the later-arrival sub-pulses having lower frequencies \citep{Hessels18,Chime19}.
This time-frequency structure is reminiscent to the Type III solar bursts and the decametric radiation from Jupiter \citep{Bastian98,Treumann06}. However, it is unclear whether the same mechanism are at work, as the FRBs 
are at cosmological distances and have extremely high brightness temperatures.
Plasma lensing may cause a sub-pulse drift, but both upward and downward drifts are expected \citep{Cordes17}. In contrast,
only the downward drifting is seen in the repeating FRBs. A mechanism intrinsic to the FRB source is most likely the origin of the drift. 
One such mechanism has been proposed in the framework of magnetar-wind-driven external shock synchrotron maser  \citep{Metzger19}. However, in this model it is not clear why such down drifting does not occur in consecutive individual bursts.

Here we propose an alternative model by invoking coherent curvature radiation in a neutron star (NS) magnetosphere. 
Sub-pulse drifting is a well-known phenomenon in radio pulsars \citep{Rankin90}, which can be interpreted as ${\bf E \times B}$ drift in the inner gap where the particles are accelerated from the polar cap region \citep{RS75}. 
Curvature radiation from charge bunches from pulsar magnetospheres has been invoked to interpret FRB coherent radio emission by several authors \citep[e.g.][]{Katz14,Kumar17,Lu18,Yang18}. 
In this letter, we propose a generic geometrical mechanism to account for the observed time-frequency downward drifting from the two repeating FRBs. The model is described in \S\ref{sec:drift}, and its applications in two specific scenarios are discussed in \S\ref{sec:models}.

\section{Geometric model}\label{sec:drift}

\begin{figure}
\includegraphics[width=0.48\textwidth]{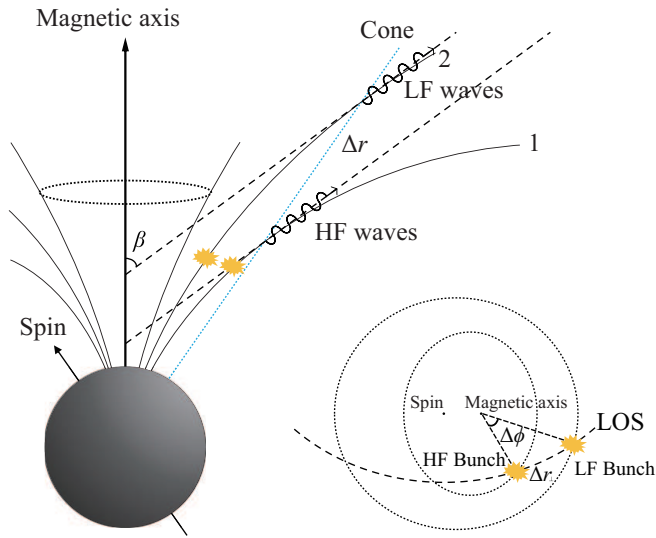}
\caption{\small{A schematic diagram of the first scenario, with sparks originating from the polar gap region. The HF waves are emitted from the lower altitudes than the LF waves. The left panel shows the initial configuration when the two sparks are produced around the same location. The dashed lines show the LOS. The second spark sweeps the LOS at a higher altitude. The right panel shows the sky map of two sparks. These two sparks sweep the LOS at different heights at different times.}}
\label{fig1}
\end{figure}

We consider a generic model of coherent curvature radiation by bunches of charged particles in a NS magnetosphere. 
The specific geometry does not matter, as long as the bunches are generated abruptly and stream outwards along open magnetic field lines. The field lines sweep across the line of sight as the magnetosphere rotates. The observer sees emission from several bunches from neighboring magnetic field lines. 
Assuming that the Lorentz factors of the bunches are the same from each other and do not evolve significantly as they stream along the field lines, the bunches observed earlier emit curvature radiation in more curved part of the field lines, and therefore have higher frequencies.
In contrast, the bunches observed later emit in less curved part of the field lines with lower frequencies.

Figure \ref{fig1} shows a schematic plot of one version of such field lines, where the ``sparks'' are produced from the inner magnetosphere of the open field lines of a NS. The sparks are produced at a low height due to a sudden release of energy, e.g. by magnetic reconnection or crust cracking. Several bunches are released around the same time and continuously radiate along neighboring field lines. In the plot, two locations are marked for the sub-pulses of high frequency (HF) and low frequency (LF). The two locations 
for the two sub-pulse emission are different in radius ($\Delta r$) and in azimuthal angle ($\Delta\phi$).

The emission frequency of curvature radiation reads $\nu = (3/4\pi) \gamma^3 (c/\rho)$, where $\rho$ is the curvature radius, and $c$ is speed of light. Assuming a constant Lorentz factor $\gamma$ of the charges, the change in the typical curvature radiation frequency is given by
\be
\Delta \nu=-\frac{3c\gamma^3\Delta\rho}{4\pi\rho^2}=-\nu\frac{\Delta \rho}{\rho},
\label{eq1}
\ee
where $\Delta\rho$ is the change in the curvature radius between the two emitting points.
Observationally, $\Delta \nu / \nu$ is of the order of 0.1 \citep{Hessels18,Chime19}, so  
we can infer $\Delta\rho/\rho\sim0.1$ for a constant $\gamma$.

If the bunch scale is smaller than the half-wavelength ($\sim10$\,cm, for 1 GHz), the phase of emission radiated by each particle in the bunch would be approximately the same, so coherent radio emission is produced \citep{Melrose17,Kumar17,Yang18}.
The GHz curvature radiation time scale for such a bunch is $1$\,ns, which is much shorter than that of the observed pulse duration $\sim1$\,ms, so there must be more than one bunch sweeping cross the line of sight (LOS) \citep{Yang18}.
Such intense sparks likely happen in an environment with abrupt release of a huge amount of energy to produce FRBs. 

Most generically, the observed time delay of LF wave with respect to the HF wave can be written as
\be
\Delta t=\Delta t_{\phi}+\Delta t_{r},
\label{eq2}
\ee
where $\Delta t_{\phi}$ is the interval between the two emission beams sweeping across the LOS, and $\Delta t_r$ is the 
delay of emission in the radial direction for the two sparks, i.e. the retardation delay (see Fig.1).  
One can generally write
\be
\Delta t_{\phi}=\frac{\Delta r_\perp}{v},
\label{tS}
\ee
where $\Delta r_\perp$ is the projected horizontal 
distance between the HF emitting region and the  LF emitting region, and $v$ is the projected 
speed of the sweeping beam.
As the two sparks are generated simultaneously but the observed emissions from the two sparks are emitted at different epochs, the delay of receiving the two signals due to the retardation delay can be estimated as
\be
\Delta t_{r}=\frac{\Delta r}{v_e}-\frac{\Delta r}{c}\approx\frac{\Delta r}{2\gamma_e^2c},
\label{tR}
\ee
where $v_e \sim c$ is the velocity of the electrons (or pairs) in the bunches, and $\gamma_e$ is its corresponding Lorentz factor. 

\section{Applications}\label{sec:models}

In this section, we apply this generic geometrical model to two specific scenarios of FRB production. The first scenario is a transient pulsar sparking model with the FRB originating from the pulsar inner gap region. The magnetic field configuration in this scenario may be approximated as a simple dipole. The second scenario is the cosmic comb model \citep{Zhang17,Zhang18}. The sparks are suddenly generated upon the interaction between the external plasma stream and the pulsar magnetosphere, which flow along the open field lines in the sheath. The field line configuration is not dipolar, but is more stretched. In both cases, the sparks propagate from high-curvature regions to low-curvature regions, leading to frequency downward drifting. We now discuss these two scenarios in turn.

\subsection{Polar gap sparking}

For the first scenario, we consider an FRB generated from the polar gap region of a pulsar. This could be related to a young regular field pulsar \citep[e.g.][]{Connor16,Cordes16} or a young magnetar with the emission coming from the inner magnetosphere \citep{Kumar17}. 

We consider a scenario similar to the polar gap sparking of the regular pulsars \citep{RS75}. However, instead of invoking regular, continuous sparks, we envisage a sudden, violent sparking process from the surface, possibly triggered by an abrupt crust cracking that leads to an abrupt magnetic field dissipation. A significant deviation from the regular magnetic field configuration is triggered, which leads to coherent curvature radiation by bunches of charged particles in a lotus of field lines \citep{Yang18}. The perturbation propagates along the field lines outwards, leading to multiple sparks emitting in adjacent field line bundles traveling with a similar Lorentz factor. 

Consider that the polar gap of the pulsar is enclosed within the last open field lines with a polar angle $\theta_p=0.1(P/10\,{\rm ms})^{-1/2}$, where $P$ is the period of the pulsar. For a dipole magnetic field, a magnetic field line can be described as
\be
u=\frac{R\sin^2\theta}{r},
\label{eq6}
\ee
where $R$ is the radius of the NS surface, and $u$ is a dimensionless constant. 
The curvature radius of the field line is (for $\theta\lesssim0.5$)
\be
\rho=\frac{r(1+3\cos^2\theta)^{3/2}}{3\sin\theta(1+\cos^2\theta)}\approx\frac{4r}{3\sin\theta}.
\label{eq7}
\ee
For $\gamma_e=300$, the curvature radius is estimated to be $\rho\simeq1.9\times10^8$\,cm to produce $\sim$ GHz curvature radiation.

For a dipolar geometry, the time for the line to sweep a phase $\Delta \phi$ is given by
\be
\Delta t_{\phi}=\frac{P\sin\beta\Delta\phi} {2\pi\sin(\alpha+\beta)},
\label{eq3}
\ee
where $P$ is the period of the pulsar, $\alpha$ is the magnetic inclination angle and $\beta$ is the impact angle of LOS with respect to the magnetic axis. 
In this scenario, $\Delta t_\phi$ only depends on the geometry of the pulsar.
As an example, we assume $\Delta r=0.01\rho$.
From equation (\ref{tR}), one can estimate the retardation time delay to be $\Delta t_{\rm r}\simeq10$\,ns, which is much smaller than the observed interval times between sub-pulses $\sim0.1-10$\,ms \citep{Hessels18,Chime19}.
Hence, the time delay of LF waves with respect to the HF waves is mainly given by the sweeping delay $\Delta t_\phi$.

Combining equations (\ref{tS}), (\ref{eq6}) and (\ref{eq7}), one gets
\be
\dot\nu=A_g\nu=\frac{2\pi\sin(\alpha+\beta)\Delta u}{uP\sin\beta\Delta\phi}\nu,
\label{eq9.1}
\ee
According to equation (\ref{eq9.1}), when the geometrical condition of $\Delta \phi/\Delta u\simeq-2\pi u^{-1}(P/10\,{\rm ms})^{-1}[\sin(\alpha+\beta)/\sin\beta]$ is satisfied, the drifting rate is very similar to what is observed in FRB 121102 \citep{Hessels18}.
If $\Delta t_\phi \ll1/A_g$, the central frequency decreases linearly with time.
This scenario matches the observations of FRB 180814.J0422+73 well \citep{Chime19}.

At the same height, electrons are in the different trajectories with essentially the same curvature radius. Since different field lines have slightly different curvatures, the condition of coherence is that the bunch opening angle $\Delta \phi_{\rm b}$ should be smaller than $1/\gamma_e$ \citep{Yang18}. Defining $\nu_{\phi}=12c/(\pi \rho\Delta\phi_{\rm b}^3)$, the condition $\nu < \nu_\phi$ can be translated to $\Delta \phi_{\rm b}<1/\gamma_e$.
Observationally, the sub-pulse interval time is of the order of milliseconds \citep{Hessels18,Chime19}, 
$\Delta t \sim (1  {\rm ms})  \Delta t_{\rm ms}$. 
The condition $\Delta \phi < 1/\gamma_e$ can be satisfied if the pulsar period satisfy $P > \gamma_e \Delta t = 0.3  {\rm s} (\gamma_e/300) \Delta t_{\rm ms}$. 

\subsection{Cosmic comb}

\begin{figure}
\includegraphics[width=0.45\textwidth]{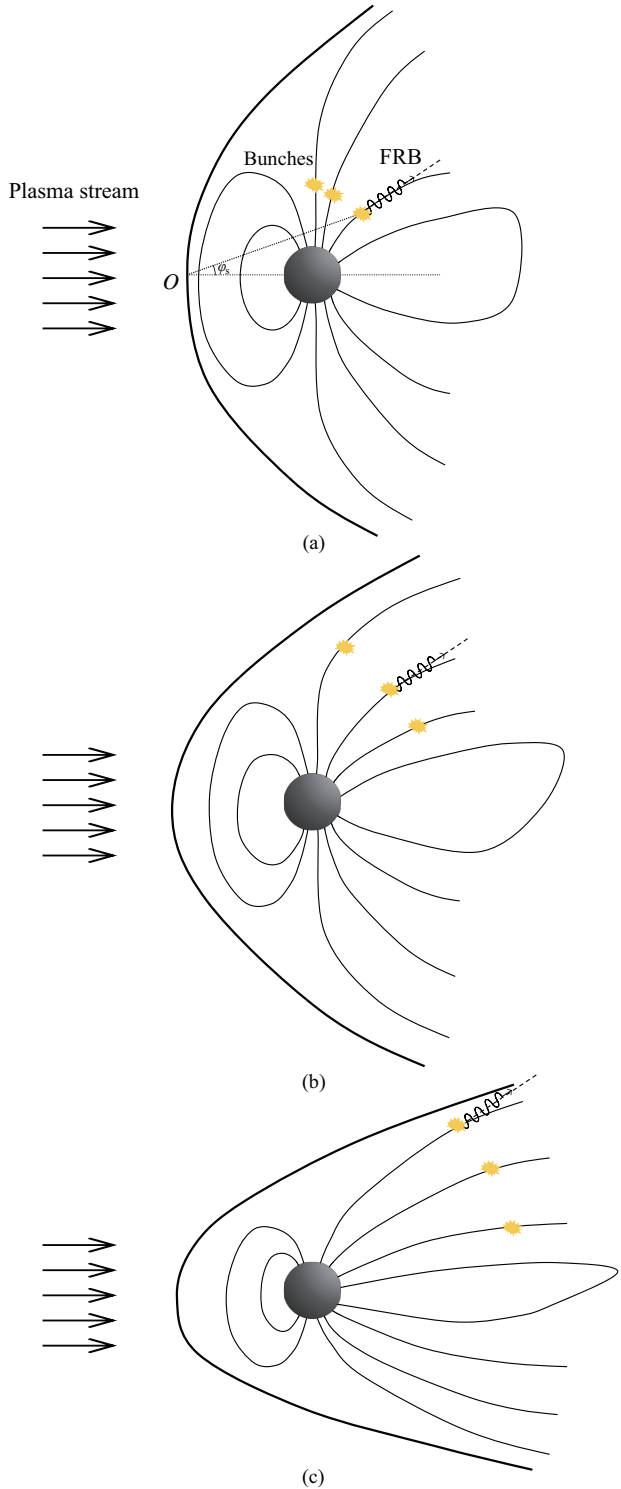}
\caption{\small{A schematic diagram of the second scenario in the cosmic comb model. The sparks are produced in the distorted sheath region which stream outwards along the field lines. For the illustrative purpose, the separations between the field lines are stretched. Sparks from different field lines sweep the LOS at different times when the sparks reach different heights. The spark observed at a later epoch emits at a less curved part of field line and thus has a lower frequency. A burst with three sub-pulses are 
shown for illustration, with three epochs: (a) the inner spark emission beams towards the LOS; (b) an intermediate spark emission beams towards the LOS; (c) the outer spark beams towards the LOS.}}
\label{fig2}
\end{figure}

In the cosmic comb model \citep{Zhang17,Zhang18} a plasma stream from a nearby source interacts with a pulsar. 
Similar to solar wind interacting with the earth magnetosphere, the external stream would re-structure the magnetosphere of the pulsar, forming an elongated magnetosphere surrounded by a sheath. The FRB is seen when the sheath plasma sweeps the LOS. 
For GHz radio waves, one requires $\gamma_e \sim 10^3$ for
the curvature radius $\rho\sim10^{10}$\,cm that matches the light cylinder radius $R_{\rm LC}=4.8\times10^9 \ {\rm cm} \ (P/1\,\rm s)$.
Since this is an abrupt process caused by the ram pressure overcoming the magnetic pressure, the field line is significantly distorted from the dipolar form.  

We envisage that the sudden distortion of the magnetosphere would drive significant electric density fluctuation with respect to the original Goldreich-Julian value, forming sparks or bunches of charged particles in a lotus of field lines around the same time. These sparks from different field lines stream outwards and sweep the LOS at different times. Figure \ref{fig2} shows a schematic view of this process for three different epochs when three sparks sweep the line of sight. One can see that the spark observed by the observer earlier originates at a lower altitude and hence has a higher frequency. As different field lines sweep across the LOS, emission with progressively decreasing frequency is observed due to the progressively larger curvature radius along these field lines. 

Again the retardation delay time $\Delta t_r \sim 1.7\times10^{-5} \ {\rm s}~ \Delta r_{12}~ \gamma_{e,3}^{-2} \ll \Delta t$ (The convention $Q_n=Q/10^n$ in cgs units is adopted). Therefore, the observed delay time is mostly defined by the sweeping delay, which reads
\be
\Delta t_{\phi}\simeq\frac{\Delta R_{s}}{v_{s}\gamma_e}\approx (3 \ {\rm ms}) \Delta R_{s,10} ~v_{s,-1}^{-1} ~\gamma_{e,3}^{-1},
\label{eq8}
\ee
where $\Delta R_{s}$ is the size of the sheath, and $r_\perp = R_{s}/\gamma$ is the projected distance in the sky when the emission beam is observed, and $v_{s}\simeq 0.1c ~v_{s,-1}$ is the velocity of the stream that combs the magnetosphere.
This is consistent with the observed millisecond interval time of the sub-pulses.
Equation (\ref{eq8}) has properties similar to equation (\ref{eq3}). 

Combining equations (\ref{eq1}), (\ref{tS}) and (\ref{eq8}), one can obtain
\be
\dot\nu=A_c\nu=-\frac{v_{\rm s}\gamma\Delta\rho}{\rho\Delta R_{\rm s}}\nu
\label{eq10}
\ee
for the cosmic comb model. One can estimate that
$\Delta \rho/\Delta R_{\rm s}\simeq0.3 \rho_{10} \gamma_{e,3}^{-1}$.
The frequency drifting rates would decrease linearly with $\nu$, which is consist with the observations of FRB 121102.
The drifting rate would be a constant when $\Delta t\ll1/A_c$ for each multi-sub-pulse sequence. In such a situation, the result matches the observations of FRB 180814.J0422+73.

\subsection{Drifting rates}

\begin{figure}
\includegraphics[width=0.45\textwidth]{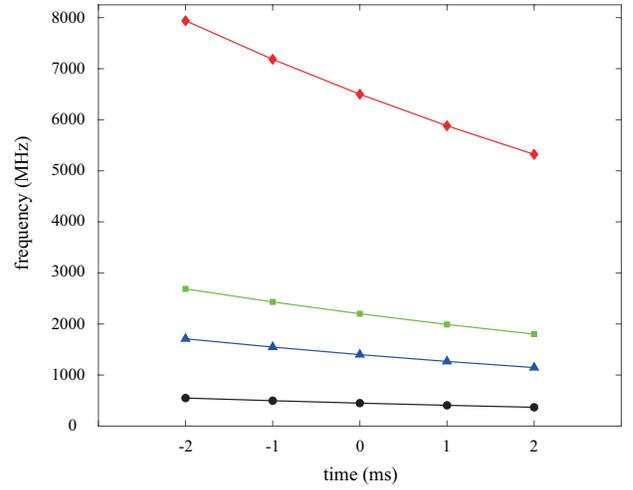}
\caption{\small{Simulated sub-burst central frequency as a function of the arrival time. We assume $A_g=A_c=-0.01\,\rm{ms^{-1}}$. The sub-burst sequences have different central frequencies with the same interval time $\Delta t=1$\,ms: 6.5 GHz (red diamonds), 2.2 GHz (green squares), 1.4 GHz (blue triangles), and 400 MHz (black dots).}}
\label{fig3}
\end{figure}

Equations (\ref{eq9.1}) and (\ref{eq10}) show that both models share the similar feature of frequency down-drifting. In Figure \ref{fig3}, we show the simulated sub-pulse central frequency drift as a function of the arrival time for the parameter $A_g=A_c=-0.01\,\rm{ms^{-1}}$. We fix $\Delta t=1$\,ms but allow the central frequency to vary. From up to down, different curves (with different colors) stand for different central frequencies: 6.5 GHz (red diamonds), 2.2 GHz (green squares), 1.4 GHz (blue triangles), and 400 MHz (black dots). These results are generally consistent with the observations of the two FRB repeaters \citep{Hessels18,Chime19}. 

\subsection{Particle cooling and acceleration}
In the above discussion, we have assumed a constant $\gamma$ for both models. For typical FRB parameters, both models involve rapid cooling of the emitting particles (the cooling rate increases by a factor of $N_e$ for coherent emission by bunches, where $N_e$ is the number of net electrons in the bunch) and therefore require continuous acceleration of the bunched particles. Very generally, the cooling timescale of curvature radiation in the observer's rest frame can be written as \citep{Kumar17}
\be
t_{\rm cool}\sim\frac{27m_{\rm e}c^3\gamma_{\rm e}^3}{16\pi^2e^2\nu^2N_{\rm e}}\sim1.8\times10^{-13} \gamma_{\rm e,2}^3\nu^{-2}_{9}(N_{\rm e,23})^{-1}\rm\,s.
\label{tc}
\ee
Therefore, to sustain a constant Lorentz factor within a lab-frame time duration of $\gamma^2/\nu$, one requires that there exists an electric field parallel $E_{\parallel}$ to the B-field that can accelerate electrons, which is given by
\be
E_{\parallel}\simeq\frac{\gamma_{\rm e}m_{\rm e}c}{(et_{\rm cool})}\sim3.1\times10^7\nu_9^2N_{\rm e, 23} \gamma_{\rm e, 2}^{-2}\,\rm esu.
\label{Ep}
\ee

For the scenario of polar gap sparking, the electron number may be described by \citep[e.g.][]{Kumar17}
\be
N_{\rm e}\simeq\frac{\mu Bc^2\gamma_{\rm e}^3}{\nu^3 eP}=1.9\times10^{24}\mu B_{14} \gamma_{\rm e,2}^3P^{-1}_{-1}\nu_9^{-3},
\ee
where $\mu$ is the normalized fluctuation of electrons deviated from the Goldreich--Julian density. The required electric field is calculated as $E_{\parallel}\sim 5.9\times10^8 \mu B_{14} P^{-1}_{-1} \gamma_{\rm e, 2}\nu_9^{-1}\,\rm esu$.
One possible mechanism to create such an electric field is the sudden magnetic reconnection in the magnetosphere.

Within the cosmic comb model, the electron number is given by \citep{Yang18}
\be
N_{\rm e}\simeq\frac{\mu \eta BR^3L}{2\pi e R_{\rm LC}^2}=3.3\times10^{19}\mu \eta B_{12}R^3_6L_1(R_{\rm LC,10})^{-2},
\ee
where $\eta R_{\rm LC}^2$ is the cross section of the bunch in nearly parallel field lines in the combed magnetosphere, and $L \sim \lambda$ is the thickness of the bunch, which is comparable to the wavelength $\lambda$ of the emission. 
The required electric field for tthis model is then $E_{\parallel} \sim 100 \mu \eta B_{12}\nu_9^2R^3_6 L_1 (R_{\rm LC,10})^{-2} \gamma_{\rm e,3}^{-2}\rm\,esu$. The strong ram pressure of the stream likely would trigger magnetic reconnection and provide the required electric field to accelerate electrons.

\section{Summary and discussion}

We proposed a generic geometrical mechanism to explain the frequency downwards drifting within the framework of coherent curvature radiation in the magnetosphere of a NS. As long as the sparks or bunches of charged particles are produced abruptly from the inner magnetosphere of a NS, and stream outwards along the open field lines, a spark observed at an earlier time 
was always emitted in a more curved part of field line, hence at a higher frequency than one observed later, which had 
traveled to a less curved part of the field line, hence emitting at a lower frequency. As a result, the frequency-time downward drifting is a natural consequence of coherent curvature radiation. We argue that this may be considered an evidence of that the FRB radio 
emission originates from a pulsar magnetosphere. We apply this generic geometrical model to explain the frequency drifting within two scenarios: a) the transient pulsar-like sparking from the inner gap of a slowly rotating NS; and b) the cosmic comb. Both models can reproduce the observations with reasonable parameters.

For the transient sparking scenario of isolated NSs, the condition is that the NS rotation period cannot be too short. This actually poses some constraints on the spindown-powered models. For the young pulsar model in supernova remnants \citep[e.g.][]{Connor16,Cordes16}, a slow rotator would give a spindown luminosity that is significantly below the FRB luminosity, making it difficult to power FRBs. For the magnetically powered (magnetar) models, the energy budget issue is less demanding \citep{Kumar17}. However, the requirement of having emission from the open field line region poses some constraints on some versions of the model \citep[e.g.][]{Lu19}. Alternatively, the FRBs may be triggered internally by, say, starquakes \citep{Wang18}. In this case, the FRBs should be accompanied by global oscillations and glitches. The cosmic comb model \citep{Zhang17,Zhang18} invokes the outer magnetosphere of a NS as the site of FRB emission. It can also naturally produces sub-pulse down-drifting, with the ultimate energy coming from the kinetic energy of the external stream. 

In our geometric model, the sparks are modeled as isolated bunches for simplicity. In reality, the outflow is likely continuous in adjacent field lines but with density fluctuations. This would give rise to continuous emission with distinct peaks, as the observations show. In contrast of the continuous sparking in the polar cap region of normal pulsars, our model invokes a sudden, violent sparking process. The FRB flow is likely abrupt and non-uniform across different field lines, which is likely the case in both scenarios discussed in the paper.

\acknowledgments
{{ We are grateful to the referee for a constructive comment and Jim Cordes, Jin-Lin Han, Jiguang Lu, Ben Margalit, Shriharsh Tendulkar, Chen Wang, Yuan-Pei Yang as well as the members in the pulsar group at Peking University for helpful discussion and comments. W.Y.W. and X.L.C. acknowledge the support of MoST Grant 2016YFE0100300, the NSFC Grants 11633004, NSFC 11473044, 11653003, and the CAS grants QYZDJ-SSW-SLH017, and CAS XDB 23040100. R.X.X. acknowledges the support of National Key R\&D Program of China (No. 2017YFA0402602), NSFC 11673002 and U1531243, and the Strategic Priority Research Program of CAS (No. XDB23010200).}}


\begin{thebibliography}{}

\bibitem[Bassa et al.(2017)]{Bassa17} Bassa, C.~G., Tendulkar, S.~P., Adams, E.~A.~K., et al.\ 2017, \apjl, 843, L8 

\bibitem[Bastian et al.(1998)]{Bastian98} Bastian, T.~S., Benz, A.~O., \& Gary, D.~E.\ 1998, \araa, 36, 131 

\bibitem[Chatterjee et al.(2017)]{Chatterjee17}Chatterjee, S., Law, C. J., Wharton, R. S., et al.\ 2017, \nat, 541, 58

\bibitem[CHIME/FRB Collaboration et al.(2019)]{Chime19} CHIME/FRB Collaboration, Amiri, M., Bandura, K., et al.\ 2019, \nat, 566, 235 

\bibitem[Connor et al.(2016)]{Connor16} Connor, L., Sievers, J., \& Pen, U.-L.\ 2016, \mnras, 458, L19

\bibitem[Cordes \& Wasserman(2016)]{Cordes16} Cordes, J.~M., \& Wasserman, I.\ 2016, \mnras, 457, 232 

\bibitem[Cordes et al.(2017)]{Cordes17} Cordes, J.~M., Wasserman, I., Hessels, J.~W.~T., et al.\ 2017, \apj, 842, 35 

\bibitem[Hessels et al.(2018)]{Hessels18} Hessels, J.~W.~T., Spitler, L.~G., Seymour, A.~D., et al.\ 2018, arXiv:1811.10748 

\bibitem[Katz(2014)]{Katz14} Katz, J.~I.\ 2014, \prd, 89, 103009 

\bibitem[Keane et al.(2012)]{Keane12}Keane, E. F., Stappers, B. W., Kramer, M., \& Lyne, A. G. 2012, \mnras, 425, L71

\bibitem[Kulkarni et al.(2014)]{Kulkarni14} Kulkarni, S.~R., Ofek, E.~O., Neill, J.~D., Zheng, Z., \& Juric, M.\ 2014, \apj, 797, 70

\bibitem[Kumar et al.(2017)]{Kumar17} Kumar, P., Lu, W., \& Bhattacharya, M.\ 2017, \mnras, 468, 2726

\bibitem[{{Lorimer} {et~al.}(2007){Lorimer}, {Bailes}, {McLaughlin}, {Narkevic}, \& {Crawford}}]{Lorimer07}{Lorimer}, D.~R., {Bailes}, M., {McLaughlin}, M.~A., {Narkevic}, D.~J., \&
  {Crawford}, F. 2007, Science, 318, 777
  
\bibitem[Lu \& Kumar(2018)]{Lu18} Lu, W., \& Kumar, P.\ 2018, \mnras, 477, 2470
  
\bibitem[Lu et al.(2019)]{Lu19} Lu, W., Kumar, P., \& Narayan, R.\ 2019, \mnras, 483, 359

\bibitem[Marcote et al.(2017)]{Marcote17} Marcote, B., Paragi, Z., Hessels, J. W. T., et al.\ 2017, \apjl, 834, L8

\bibitem[Melrose(2017)]{Melrose17} Melrose, D.~B.\ 2017, Reviews of Modern Plasma Physics, 1, 5 

\bibitem[Metzger et al.(2019)]{Metzger19} Metzger, B.~D., Margalit, B., \& Sironi, L.\ 2019, arXiv:1902.01866 
  
\bibitem[Michilli et al.(2018)]{Michilli18}  Michilli, D., Seymour, A., Hessels, J. W. T., et al.\ 2018, \nat, 553, 182
  
\bibitem[Petroff et al.(2015)]{Petroff15} Petroff, E., Johnston, S., Keane, E. F., et al.\ 2015, \mnras, 454, 457

\bibitem[Petroff et al.(2016)]{Petroff16}  Petroff, E., Barr, E. D., Jameson, A., et al.\ 2016, PASA, 33, e045

\bibitem[Rankin(1990)]{Rankin90} Rankin, J.~M.\ 1990, \apj, 352, 247 

\bibitem[Ruderman \& Sutherland(1975)]{RS75} Ruderman, M. A., \& Sutherland, P. G. 1975, \apj, 196, 51

\bibitem[Spitler et al.(2016)]{Spitler16} Spitler, L.~G., Scholz, P., Hessels, J.~W.~T., et al.\ 2016, \nat, 531, 202 

\bibitem[Tendulkar et al.(2017)]{Tendulkar17} Tendulkar, S.~P., Bassa, C.~G., Cordes, J.~M., et al.\ 2017, \apjl, 834, L7 

\bibitem[Thornton et al.(2013)]{Thornton13}Thornton, D., Stappers, B., Bailes, M., et al. 2013, Science, 341, 53

\bibitem[Treumann(2006)]{Treumann06} Treumann, R.~A.\ 2006, \aapr, 13, 229 

\bibitem[Wang et al.(2018)]{Wang18} Wang, W., Luo, R., Yue, H., et al.\ 2018, \apj, 852, 140 

\bibitem[Yang \& Zhang(2018)]{Yang18} Yang, Y.-P., \& Zhang, B.\ 2018, \apj, 868, 31 

\bibitem[Zhang(2017)]{Zhang17} Zhang, B.\ 2017, \apjl, 836, L32 

\bibitem[Zhang(2018)]{Zhang18} Zhang, B.\ 2018, \apjl, 854, L21 


\end{thebibliography}
\end{document}